\documentclass[aps,prd,groupedaddress,showpacs]{revtex4}

\usepackage{graphicx,dcolumn,bm,amssymb,amsmath,latexsym,amsfonts}

\begin{document}
\newcommand{\nonu}{\nonumber}
\newcommand{\sm}{\small}
\newcommand{\noi}{\noindent}
\newcommand{\nl}{\newline}
\newcommand{\bp}{\begin{picture}}
\newcommand{\ep}{\end{picture}}
\newcommand{\bc}{\begin{center}}
\newcommand{\ec}{\end{center}}
\newcommand{\be}{\begin{equation}}
\newcommand{\ee}{\end{equation}}
\newcommand{\beal}{\begin{align}}
\newcommand{\eeal}{\end{align}}
\newcommand{\bea}{\begin{eqnarray}}
\newcommand{\eea}{\end{eqnarray}}
\newcommand{\bnabla}{\mbox{\boldmath $\nabla$}}
\newcommand{\univec}{\textbf{a}}
\newcommand{\VectorA}{\textbf{A}}
\newcommand{\Pint}

\title{Exact solution for a binary system of unequal counter--rotating black holes}
\author{I. Cabrera-Munguia$^{1,2,}$\footnote{cabrera@zarm.uni-bremen.de}, Claus L\"ammerzahl$^{1,}$\footnote{laemmerzahl@zarm.uni-bremen.de} and Alfredo Mac\'{\i}as$^{2,}$\footnote{amac@xanum.uam.mx}}
\affiliation{$^{1}$ZARM, Universit\"{a}t Bremen, Am Fallturm, D-28359 Bremen, Germany\\
$^{2}$Departamento de F\'isica, Universidad Aut\'onoma Metropolitana-Iztapalapa A.P. 55-534, M\'exico D.F. 09340, M\'exico}


\begin{abstract}
A complete solution describing a binary system constituted by two unequal counter-rotating black holes with a massless strut in between is presented. It is expressed in terms of four arbitrary parameters: the half length of the two rods representing the black hole horizons $\sigma_{1}$ and $\sigma_{2}$, the total mass $M$ and the relative distance $R$ between the centers of the horizons. The explicit parametrization of this solution in terms of physical parameters, i.e., the Komar masses $M_{1}$ and $M_{2}$, the Komar angular momenta $J_{1}$ and $J_{2}$ (having $J_{1}$ and $J_{2}$ opposite signs) and the coordinate distance $R$, led us to a 4-parameter subclass in which the five physical parameters satisfy a simple algebraic relation, which generalizes the two statements made by Bonnor, in order to remove the additional contributions from the massless spinning rods outside the black holes. Moreover, the interaction force turns out to be of the same form as in the double-Schwarzschild static case.
\end{abstract}

\maketitle

\section{INTRODUCTION}
The equilibrium configurations of the famous double-Kerr-NUT (DKN) solution \cite{KramerNeugebauer} have been extensively studied in the last three decades. Applying regularity conditions on the symmetry axis, the balance equations were first derived by Kihara and Tomimatsu \cite{KiharaTomi,TomiKihara}. Hoenselaers solved them analytically for the case of subextreme sources, i.e., non-degenerate black holes, where a further analysis revealed the existence of ring singularities off the axis \cite{CHoenselaers}, due mainly to the fact that at least one of the Komar masses is negative \cite{Komar}. Recently, Neugebauer and Hennig \cite{HennigNeugebauer} have shown the non-existence of regular solutions describing equilibrium configurations between two rotating black holes by using the analytical solution presented by Manko \emph{et al} \cite{MRS}. Additionally, if the parameters do not satisfy the regularity conditions on the symmetry axis, there arise two kinds of singularities on the axis, known after Bonnor \cite{Bonnor1} as \emph{torsion singularity} and \emph{stress singularity}. The first one generates a region with closed timelike curves due to the presence of NUT sources, which leads to finite and semi--infinite singularities along the axis, breaking the asymptotic flatness of the solution \cite{NUT,Bonnor0}. The second one represents a strut, a conical singularity \cite{Israel}, which helps us to understand the interaction force between the two bodies by means of the gravitational attraction and the spin-spin interaction.

In the aforementioned equilibrium problem, in which one notices the absence of a strut, one always starts by solving the \emph{balance condition}, then the corresponding algebraic variables are substituted into the \emph{axis condition}. Nevertheless, one might choose the opposite way and first solve the equations for avoiding the NUT sources, with the purpose of calculating the massless strut and determine the interaction force between the two black holes. This last approach is more general and complicated to analyze than the equilibrium situation. Nowadays the 5-parameter subclass of the well-known DKN solution \cite{KramerNeugebauer} and its analytical representation in terms of independent physical parameters still remain as an open problem.\nl

One of the first attempts to describe the physical properties of two rotating black holes, was made by
Varzugin \cite{Varzugin}. He solved the corresponding Riemann-Hilbert problem, in which the {\em irreducible mass} $\sigma_{i}$ is defined as the half length of the rod representing the event horizon of the $i$--th black hole located on the symmetry axis. First, the axis conditions were formulated, with the black holes separated by a massless strut in between, in order to prevent the falling onto each other, due to the gravitational attraction.

Since, by means of the Smarr mass formula \cite{Smarr}, the irreducible mass can be related with the surface gravity and area of the black hole horizon, one could parametrize it in terms of physical Komar parameters. Varzugin found the simplest analytical solution describing a binary system constituted by identical counter-rotating black holes, where the interaction force turns out to be of the same form as in the static Schwarzschild case. The corresponding unique $\sigma$ is described by only three parameters: the Komar mass $m$ and angular momentum $j$ for the upper black hole \cite{Komar} (the lower one has the parameters $m$ and $-j$), and the coordinate distance $R$ between both constituents.

Later on, Manko \emph{et al} \cite{MRRS} used this explicit solution to provide explicit expressions for the Ernst potential and the metric in the whole space-time. This solution is equatorially antisymmetric \cite{EMR}. The axis condition is straightforwardly fulfilled and the total angular momentum of the system vanishes, i.e., $J=0$.

For the case of two identical bodies, Bonnor \cite{Bonnor1, Bonnor2} advanced two additional conditions to be satisfied in order to remove the additional contribution provided by the massless spinning rods outside the sources:
\be  (i)\,\, \frac{J_{1}}{M_{1}^{2}}+\frac{J_{2}}{M_{2}^{2}}=0, \qquad (ii)\,\, \frac{J_{1}}{M_{1}} + \frac{J_{2}}{M_{2}}=0\, . \label{Bonnorconditions}\ee

The first condition of equation (\ref{Bonnorconditions}) avoids the semi-infinite massless spinning rods located in the upper and lower parts of the symmetry axis, while the second one avoids the massless spinning rod of finite length between the two bodies.\nl

It is worthwhile to stress the fact that the most satisfactory solution describing a system of two unequal counter-rotating black holes separated by a massless strut must be characterized by five physical parameters, i.e., the Komar masses $M_{1}$, $M_{2}$ of each constituent, their respective Komar angular momenta $J_{1}$, $J_{2}$ ($J_{1}$ and $J_{2}$ having opposite sign), and the relative coordinate distance $R$ between the centers of the black hole horizons. The main difficulty to accomplish this endeavor is the problem of avoiding the NUT sources in order to be able to provide the explicit form of $\sigma_{1}$ and $\sigma_{2}$ in terms of Komar physical parameters.\nl

In this work,  by means of the Sibgatullin's method \cite{Sibgatullin,RMJ}, we first derive the extended version of the DKN solution \cite{KramerNeugebauer}. Then, we solve the axis condition for the particular case of two unequal counter-rotating black holes with a massless strut in between. We write the solution in terms of $\sigma_{1}$ and $\sigma_{2}$ as a 4-parameter subclass of the DKN problem. Later on, we calculate $\sigma_{1}$ and $\sigma_{2}$ by using the Komar integrals for the masses $M_{i}$ and the angular momenta $J_{i}$. We show that the interaction force between the black holes, provided by the strut, reduces to the same form as the one for the static double-Schwarzschild case. Moreover, the five parameters satisfy an algebraic relation, which generalizes the two statements advanced by Bonnor \cite{Bonnor1, Bonnor2} in order to remove the additional contributions made by the massless spinning rods outside of the sources, i.e., Eq.(\ref{Bonnorconditions}). At some particular value of the distance, a ``dynamic scenario" between the black holes arises, since the physical properties of one body are affected by the presence of the other one. In this description, the total angular momentum of the system is \emph{exactly} $J=J_{1}+J_{2}$. Notice that it contains only the contributions from the two sources.\nl

The outline of the paper is as follows. In Sec. II, all the necessary elements to construct the DKN solution are presented. In Sec. III, a 4-parametric exact solution describing two counter-rotating black holes separated by a massless strut is analyzed. In Sec. IV, the parametrization of the solution in terms of the physical Komar parameters is accomplished. In Sec. V the concluding remarks are presented.

\section{THE DOUBLE-KERR-NUT SOLUTION}
The Papapetrou line element describing stationary axisymmetric space-time reads \cite{Papapetrou}
\be ds^{2}=f^{-1}\left[e^{2\gamma}(d\rho^{2}+dz^{2})+\rho^{2}d\varphi^{2}\right]- f(dt-\omega d\varphi)^{2},
\label{Papapetrou}\ee

\noi where $f$, $\omega$ and $\gamma$ are unknown functions depending only on the cylindrical coordinates $(\rho,z)$. According to the Ernst formalism \cite{Ernst}, the vacuum Einstein field equations for these particular stationary axisymmetric space-times read
\be ({\rm{Re}}{\cal{E}}) \Delta {\cal{E}}=\bnabla{\cal{E}} \cdot \bnabla{\cal{E}}, \label{vacuum}\ee

\noi where \bnabla and $\Delta$ denote the gradient and Laplace operators, respectively, expressed in cylindrical coordinates and acting over the complex Ernst potential ${\cal{E}}=f+ i\Psi$. For any solution of equation (\ref{vacuum}), the metric functions $\omega$ and $\gamma$ of the line element (\ref{Papapetrou}) can be obtained from the following system of differential equations:
\bea \omega_{\rho} &=& -\rho f^{-2}\Psi_{z},\qquad \qquad \quad\,\, \,\, \,\, \omega_{z}=\rho f^{-2}\Psi_{\rho},\\
4\gamma_{\rho}&=&\rho f^{-2} \left(|{\cal{E}}_{\rho}|^{2}-|{\cal{E}}_{z}|^{2}\right),\, \, \, \,\, \,\,
2\gamma_{z}=\rho f^{-2} \rm{Re}({\cal{E}}_{\rho}\,{\bar{\cal{E}}}_{z}),
\label{metricfunctions}\eea

\noi where the bar over a symbol represents the complex conjugate operation, $|x|^{2}=x \bar{x}$, and the subscripts $\rho$ or $z$ denote partial differentiation. In order to solve the non-linear equation (\ref{vacuum}), we will use the powerful mathematical technique based on the soliton theory known as Sibgatullin's method \cite{Sibgatullin}. The extended DKN problem \cite{KramerNeugebauer} can be constructed easily by applying this method as it is done in \cite{RMJ} for the case of two bodies without electromagnetic field ($\Phi=0$). Let us start by writing the Ernst potential on the symmetry axis as follows:
\be {\cal{E}}(\rho=0,z)\equiv e(z) = 1 + \frac{e_{1}}{z-\beta_{1}} + \frac{e_{2}}{z-\beta_{2}}.
\label{Ernstsymmetryaxis}\ee

The set of complex constant parameters $\{e_{k},\beta_{k}\}$ consists of eight real parameters related with the Simon's multipolar moments \cite{simon}. Once we know the value of the Ernst potential on the symmetry axis, the complex potential in the whole space can be obtained from the Sibgatullin's integral
\be {\cal{E}}(\rho,z)=\frac{1}{\pi}\int_{-1}^{1}\frac{\mu(\zeta)e(\xi)d\zeta}{\sqrt{1-\zeta^{2}}}, \ee

\noi whose unknown function $\mu(\zeta)$ satisfies an integral equation
\be -\hspace{-0.37cm}\int_{-1}^{1}\frac{\mu(\zeta)[e(\xi)+\widetilde{e}(\eta)]
d\zeta} {(\xi-\eta)\sqrt{1-\zeta^{2}}}=0, \label{integralequation}\ee

\noi and a normalization condition
\be \frac{1}{\pi}\int_{-1}^{1}\frac{\mu(\zeta)d\zeta}{\sqrt{1-\zeta^{2}}}=1,\label{normalization}\ee

\noi where $ \widetilde{e}(\eta)\equiv \overline{e(\bar{\eta})}$, and $ -\hspace{-0.30cm}\int$ is representing
a principal value integral. In addition, $e(\xi)$ is the local holomorphic continuation of $e(z)$ on the complex
plane $z+i\rho$, with $\xi=z+i\rho\zeta$,\, $\eta=z+i\rho\tau, \,\forall \,\zeta,\tau\in[-1,1]$. Since $e(z)$ is a rational function, the corresponding function $\mu(\zeta)$ can be assumed of the same polynomial form:
\be\mu(\zeta)=A_{0}+\sum_{n=1}^{4}A_{n}(\xi-\alpha_{n})^{-1},\ee

\begin{figure}[ht]
\centering
\includegraphics[width=2.0cm,height=6.0cm]{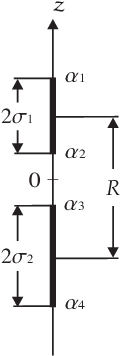}
\caption{Location of two unequal Kerr black holes on the symmetry axis represented by the rods of length $2\sigma_{1}$ and $2\sigma_{2}$, where the roots satisfy the condition $\sum \alpha_{n}=0$. The two bodies are disconnected if the axis condition and the relation $R>\sigma_{1} + \sigma_{2}$ are satisfied.}
\label{DKRR}\end{figure}

\noi where the coefficients $A_{0}$ and $A_{n}$ are determined by means of Eqs.(\ref{integralequation})-(\ref{normalization}), and the constants $\alpha_{n}$ represent the location of the sources on the symmetry axis, they are the roots of the following characteristic equation, see Fig. \ref{DKRR},
\be e(z) + \tilde{e}(z)=0.\label{characteristic}\ee

Replacing Eq.(\ref{Ernstsymmetryaxis}) into Eq.(\ref{characteristic}), it is possible to show that the old parameters $\{e_{k},\beta_{k}\}$ and the new ones $\{\alpha_{n},\beta_{k}\}$ are related through the following relations:
\be e_{1}=\frac{2(\beta_{1}-\alpha_{1})(\beta_{1}-\alpha_{2})(\beta_{1}-\alpha_{3})(\beta_{1}-\alpha_{4})}
{(\beta_{1}-\beta_{2})(\beta_{1}-\bar{\beta}_{1})(\beta_{1}-\bar{\beta}_{2})}, \qquad
e_{2}=\frac{2(\beta_{2}-\alpha_{1})(\beta_{2}-\alpha_{2})(\beta_{2}-\alpha_{3})(\beta_{2}-\alpha_{4})}
{(\beta_{2}-\beta_{1})(\beta_{2}-\bar{\beta}_{1})(\beta_{2}-\bar{\beta}_{2})}\cdot\label{the-es} \ee

After tedious but straightforward calculations, the solution describing the extended DKN problem can be obtained. The Ernst potential ${\cal{E}}$ and the corresponding metric functions $f$, $\omega$ and $\gamma$ can be written in the following explicit form \cite{MankoRuiz}:
\bea \begin{split} {\cal{E}}&=\frac{E_{+}}{E_{-}},\qquad f=\frac{E_{+}\bar{E}_{-}+\bar{E}_{+}{E}_{-}}
{2 |E_{-}|^{2}}, \qquad \omega=-\frac{ {4\rm {Im}}\left[\bar{E}_{-}G\right]}
{E_{+}\bar{E}_{-}+\bar{E}_{+}{E}_{-}},\qquad e^{2\gamma}=\frac{E_{+}\bar{E}_{-}+\bar{E}_{+}{E}_{-}}{2|K_{0}|^{2} \prod_{n=1}^{4}r_n}, \\
E_{\pm}&=\left|
\begin{array}{ccccc}
1 & 1  & 1 & 1 & 1 \\
\pm1 & {} & {}& {} & {} \\
\pm1 & & {} \mathcal{M} & {} & {}\\
0 & {} & {} & {} & {} \\
0 & {} & {} & {} & {} \\
\end{array}
\right|,\qquad
G= \left|
\begin{array}{ccccc}
0 & p_{1} & p_{2} & p_{3} & p_{4} \\
1 & {} & {} & {} & {}  \\
1 & {} & \mathcal{M}  \\
0 & {} & {} & {} & {} \\
0 & {} & {} & {} & {} \\
\end{array}
\right|,
\qquad \mathcal{M}=\left(\begin{array}{cccc}
\gamma_{11}r_{1} & \gamma_{12}r_{2} & \gamma_{13}r_{3}& \gamma_{14}r_{4} \\
\gamma_{21}r_1 & \gamma_{22}r_{2} & \gamma_{23}r_{3}& \gamma_{24}r_{4} \\
\kappa_{11} & \kappa_{12} & \kappa_{13}& \kappa_{14}\\
\kappa_{21} & \kappa_{22} & \kappa_{23}& \kappa_{24}\\
\end{array}
\right),\\
 K_{0}&=\left|
\begin{array}{cccc}
\gamma_{11} & \gamma_{12} & \gamma_{13}& \gamma_{14} \\
\gamma_{21} & \gamma_{22} & \gamma_{23}& \gamma_{24} \\
\kappa_{11} & \kappa_{12} & \kappa_{13}& \kappa_{14}\\
\kappa_{21} & \kappa_{22} & \kappa_{23}& \kappa_{24}\\
\end{array}
\right|, \qquad
p_{n}=z-\alpha_{n}-r_{n}, \qquad \gamma_{kn}=(\alpha_{n}-\beta_{k})^{-1}, \qquad \kappa_{kn}=(\alpha_{n}-\bar{\beta}_{k})^{-1}, \qquad\\
r_{n}&=\sqrt{\rho^{2}+(z-\alpha_{n})^{2} }. \end{split} \label{generalsolution}\eea

These last expressions represent a solution parameterized by eight parameters, $\alpha_{n}$, $n=1,2,3,4$, and $\beta_{k}$, $k=1,2$. However, when $\alpha_{n}$ are set to be real parameters, the solution describes a binary system constituted by two Kerr black holes, where the two horizons are defined on the symmetry axis by the intervals $\alpha_{1}\geq z \geq  \alpha_{2}$ and $\alpha_{3}\geq z \geq  \alpha_{4}$. It is important to note that the above solution was constructed assuming asymptotic flatness at spatial infinity, where $f\rightarrow1$, $\gamma\rightarrow 0$ and $\omega\rightarrow0$ (in the absence of NUT sources), the metric functions $\gamma$ and $\omega$ automatically fulfil the following conditions on the symmetry axis: $\gamma(\alpha_{1}<z<\infty)=\gamma(-\infty<z<\alpha_{4})=0$, and $\omega(\alpha_{1}<z<\infty)=0$, thus establishing an \emph{elementary flatness} on the upper part of the symmetry axis.

\section{THE FOUR-PARAMETER SUBCLASS}
For the case in which the binary system is located on the symmetry axis in such way that the roots $\alpha_{n}$ satisfy the condition $\sum\alpha_{n}=0$, only seven parameters are needed to characterize such solution.
In order to get rid of the NUT sources between the objects in the lower part of the symmetry axis, i.e., regions
$ \alpha_{3}<z<\alpha_{2}$ and $-\infty<z<\alpha_{4}$, thus regularizing the symmetry axis outside the
sources, and determining the solution for two counter-rotating black holes with a massless strut in between, a well-known conical line singularity \cite{Israel}, we must impose the following two
conditions on the metric function $\omega$:
\be \omega(\rho=0,\alpha_{2}<z<\alpha_{3}) = 0, \qquad  \omega(\rho=0,-\infty<z<\alpha_{4}) = 0.  \label{twoconditions}\ee

We note that the second condition in Eqs.(\ref{twoconditions}) implies the vanishing of the gravitomagnetic
monopole (NUT parameter \cite{NUT}), which also can be determined asymptotically by the Ernst potential on the symmetry axis Eq.(\ref{Ernstsymmetryaxis}) as follows:
\be  {\rm{Im}}[e_{1}+e_{2}]=0, \label{NUTvanishing}\ee

\noi where $e_{1}$ and $e_{2}$ are defined in Eq.(\ref{the-es}). A straightforward simplification over these
two conditions in Eq.(\ref{twoconditions}) leads us to the following compact set of algebraic equations
\bea \begin{split} {\rm{Im}}\left[ \left|\begin{array}{ccccc}
0 &     1       &        1     &       1    &      1 \\
1 &\gamma_{11}  & \gamma_{12}  & \gamma_{13}& \gamma_{14}  \\
1 & \gamma_{21} & \gamma_{22} & \gamma_{23} & \gamma_{24} \\
0 & \kappa_{11} & \kappa_{12} & \kappa_{13} & \kappa_{14}\\
0 & \kappa_{21} & \kappa_{22} & \kappa_{23} & \kappa_{24}\\
\end{array}
\right| \right]&=0, \qquad
{\rm{Im}}\left[ \left|\begin{array}{ccccc}
0 &     1        &        1     &       0    &      0 \\
1 &-\gamma_{11}  & -\gamma_{12}  & \gamma_{13}& \gamma_{14}  \\
1 &-\gamma_{21}  & -\gamma_{22} & \gamma_{23} & \gamma_{24} \\
0 & \kappa_{11}  & \kappa_{12} & \kappa_{13} & \kappa_{14}\\
0 & \kappa_{21}  & \kappa_{22} & \kappa_{23} & \kappa_{24}\\
\end{array}
\right| \right]&=0. \end{split}\label{omegasregions}\eea
These last two equations reduce the seven parametric solution to a five parametric one and the complete metric can be written in a similar form as the one for the double Reissner--Nordstr\"{o}m problem \cite{MankoDRN,AB}. We will further restrict our solution to a four parametric subclass. Since the solution Eq.(\ref{generalsolution}) involves real constants $\alpha_{n}$, which determine the location of the two Kerr black hole sources on the symmetry axis, we re-parametrize them as follows
\be\alpha_{1}=\frac{R}{2}+\sigma_{1}, \qquad \alpha_{2}=\frac{R}{2}-\sigma_{1}, \quad
\alpha_{3}=-\frac{R}{2}+\sigma_{2},\qquad \alpha_{4}=-\frac{R}{2}-\sigma_{2},\ee

\noi where, as mentioned above, $\sigma_{1}$ and $\sigma_{2}$ describe the half length of the two rods representing
the black hole horizons and $R$ is the relative distance between the two centers, as it is shown in Fig. \ref{DKRR}. The lengths $\sigma_{1}$ and $\sigma_{2}$ can be written in terms of the Komar physical parameters, i.e., the individual masses $M_{1}$ and $M_{2}$, the angular momenta $J_{1}$ and $J_{2}$ and the coordinate distance $R$. In our case, these five parameters should satisfy an additional relationship, a generalization of the Bonnor's conditions in order to remove the additional contributions made by the massless spinning rods outside of the sources. The Komar integrals for the individual masses $M_{i}$ and angular momenta $J_{i}$ can be calculated through the Tomimatsu's formulae \cite{Tomimatsu0}:
\be M_{i}=-\frac{1}{8\pi}\int_{H_{i}} \omega \Psi_{z} d\varphi dz, \qquad J_{i}=-\frac{1}{8\pi}\int_{H_{i}}\omega\left(1+\frac{1}{2}\omega \Psi_{z} \right) d\varphi dz.\label{Tomi}\ee

The integrals are over the black hole horizons $H_{i}=\{\alpha_{2i}\leq z \leq \alpha_{2i-1},\,
0 \leq \varphi \leq 2\pi,\, \rho\rightarrow 0\}$, $i=1,2$. Moreover, once the conditions established in Eq.(\ref{twoconditions}), for regularizing the symmetry axis, are fulfilled, the total mass $M$ can be considered as
the sum of the individual masses $M_{1}$ and $M_{2}$. In order to solve the system of Eqs.(\ref{omegasregions}), we  obtain the total mass $M$ of the system by employing the Fodor-Hoenselaers-Perj\'es (FHP) procedure \cite{FHP} for the calculation of the Geroch-Hansen multipole moments \cite{Geroch,Hansen}, the obtained result reads:
\be  {\rm{Re}}[e_{1}+e_{2}]=-2 M. \label{totalmass}\ee

Replacing Eq.(\ref{the-es}) into Eq.(\ref{totalmass}) yields the equation
\be \beta_{1} + \beta_{2} + \bar{\beta}_{1} + \bar{\beta}_{2}=-2 M,\ee

\noi implying several possibilities on the relations between the $\beta$-parameters and the total mass $M$. The simplest choice describing the unequal counter-rotating case is the relation $\beta_{1}+ \beta_{2}=-M$. A simple calculation leads us to the following result
\bea \begin{split} \beta_{1,2} &=\frac{-M\pm\sqrt{p+iq}}{2}, \\
 p&= R^{2}-M^{2}+2\left(\epsilon_{1} - \frac{\epsilon_{2}R}{M}\right),  \\
 q&=\frac{2\sqrt{(R^{2}-M^{2})(M^{2}R^{2}-\epsilon_{2}^{2})(M^{4}-2\epsilon_{1}M^{2}+ \epsilon_{2}^{2})}}
{M(M R+\epsilon_{2})},\\
\epsilon_{1,2}&:=\sigma^{2}_{1}\pm\sigma^{2}_{2}, \end{split}\eea

\noi where the subindexes 1 and 2 are associated with $+$ and $-$ signs, respectively. Therefore, writing the Ernst potential and the metric functions in terms of the parameters $M$, $R$, $\sigma_{1}$ and $\sigma_{2}$ leads to the following 4-parametric solution for two unequal counter-rotating black holes:
\bea {\cal{E}}&=&\frac{\Lambda+2\Gamma}{\Lambda-2\Gamma},\qquad
f=\frac{|\Lambda|^{2}-4|\Gamma|^{2}}{|\Lambda-2\Gamma|^{2}}, \qquad \omega=-\frac{2 {\rm{Im}}\left[(\bar{\Lambda}-
2\bar{\Gamma})G\right]}{|\Lambda|^{2}-4|\Gamma|^{2}},\qquad
e^{2\gamma}=\frac{|\Lambda|^{2}-4|\Gamma|^{2}}{256\sigma^{2}_{1}\sigma^{2}_{2}
(M^{2}R^{2}-\epsilon^{2}_{2})^{2} r_{1}r_{2}r_{3}r_{4}},  \nonu\\
\Lambda&=&4\sigma_{1}\sigma_{2}\left(M^{4}-\epsilon^{2}_{2}\right)(r_{1}r_{2}+r_{3}r_{4})
+ \left[M^{4}R^{2}+\epsilon^{2}_{2}(R^{2}-2M^{2})\right]\left(r_{1}-r_{2}\right)(r_{3}-r_{4})+
(M^{4}-2M^{2}R^{2}+\epsilon^{2}_{2}) \nonu\\
&\times&  \left[\epsilon_{1}(r_{1}-r_{2})(r_{3}-r_{4})- 2\sigma_{1}\sigma_{2}(r_{1}+r_{2})
(r_{3}+r_{4})\right]- 2i \delta [\sigma_{1}\left(r_{1}+r_{2}\right)(r_{3}-r_{4})-
\sigma_{2}\left(r_{1}-r_{2}\right)(r_{3}+r_{4})],  \nonu\\
\Gamma&=&\sigma_{1}\left(M+\epsilon_{2}/M\right) \left[ 2\sigma_{2}M^{2}
\left(\epsilon_{2}-R^{2}\right)(r_{3}+r_{4})- R(M^{2}-\epsilon_{2})^{2}(r_{3}-r_{4})\right]\nonu\\
&-&\sigma_{2}\left(M-\epsilon_{2}/M\right) \left[ 2\sigma_{1} M^{2}\left(\epsilon_{2}+R^{2}\right)(r_{1}+r_{2})-
R(M^{2}+\epsilon_{2})^{2}(r_{1}-r_{2}) \right] \nonu\\
&+&i \delta \left[\sigma_{1}\left(M+\epsilon_{2}/M\right)(r_{3}-r_{4})-
\sigma_{2}(M-\epsilon_{2}/M)(r_{1}-r_{2}) \right], \nonu\\
G&=& 2z\Gamma + 4\sigma_{1}\sigma_{2}R \left(M^{4}-\epsilon^{2}_{2}\right)
(r_{1}r_{2}-r_{3}r_{4})+ \sigma_{1}\left(R^{2}+\epsilon_{2}\right)
(M^{2}-\epsilon_{2})^{2}(r_{1}+r_{2})(r_{3}-r_{4})\label{4parametricsolution}\\
&+& \sigma_{2}\left(R^{2}-\epsilon_{2}\right)
(M^{2}+\epsilon_{2})^{2}(r_{1}-r_{2})(r_{3}+r_{4})-2i\epsilon_{2} \delta (r_{1}-r_{2})(r_{3}-r_{4}) - \sigma_{1}\left(M+\epsilon_{2}/M\right) \nonu\\
&\times& \{ 2\sigma_{2}R \left[\epsilon^{2}_{2}+M^{2}(M^{2}-R^{2}-\epsilon_{2})\right]
(r_{3}+r_{4}) +\left[ 2M^{2}\epsilon_{1}(R^{2}-\epsilon_{2})+(2M^{2}-R^{2})\epsilon^{2}_{2}-M^{4}R^{2}
\right](r_{3}-r_{4})  \nonu\\
&+& i \delta  \left[R\left(r_{3}-r_{4}\right)-2\sigma_{2} \left(r_{3}+r_{4}\right)\right]\}
+\sigma_{2}\left(M-\epsilon_{2}/M\right)\{ 2\sigma_{1}R \left[\epsilon^{2}_{2}+M^{2}
(M^{2}-R^{2}+\epsilon_{2})\right](r_{1}+r_{2}) \nonu\\
&-&\left[ 2M^{2}\epsilon_{1}(R^{2}+\epsilon_{2})+(2M^{2}-R^{2})\epsilon^{2}_{2}-M^{4}R^{2}
\right](r_{1}-r_{2}) - i \delta \left[R\left(r_{1}-r_{2}\right)+2\sigma_{1} \left(r_{1}+r_{2}\right)\right]\},\nonu\\
\delta&:=&\sqrt{(R^{2}-M^{2})(M^{2}R^{2}-\epsilon^{2}_{2}) (M^{4}-2\epsilon_{1}M^{2}+\epsilon^{2}_{2})},\nonu
\eea

\noi where $r_{n}$ can be written in the following parameterized form:
\be r_{1,2}=\sqrt{\rho^{2}+\left(z-\frac{1}{2}R \mp \sigma_{1}\right)^{2}}, \qquad
r_{3,4}=\sqrt{\rho^{2}+\left(z+\frac{1}{2}R \mp \sigma_{2}\right)^{2}},
 \label{thedistances}\ee

\noi the indices 1,3 and 2,4 run over $+$ and $-$ signs, respectively. Obviously
solution (\ref{4parametricsolution}) has not the equatorial antisymmetry property in the sense of \cite{EMR}, the antisymmetry appears only for the case where both constituents are equal.

It is interesting to note that under the transformation $1\leftrightarrow 2$, $z\rightarrow-z$, which exchange
the physical properties and the position of the constituents, will only change the global sign of the metric function $\omega$. The corresponding Ernst potential on the symmetry axis now reads:
\bea \begin{split} e(z)&=\frac{e_{+}}{e_{-}}, \\
e_{\pm}&= z^{2} \mp M z + \frac{2M^{3}-MR^{2}-2M\epsilon_{1}\mp 2\epsilon_{2}R}{4M}-i\frac{\sqrt{(R^{2}-M^{2})
(M^{2}R^{2}-\epsilon_{2}^{2})(M^{4}-2\epsilon_{1}M^{2}+ \epsilon_{2}^{2})}}{2M(M R \mp \epsilon_{2})}. \end{split}\label{ernstsymmetry} \eea

After using the FHP procedure the total angular momentum of the system is given by
\be J= \frac{\epsilon_{2}}{2M}\sqrt{\frac{(R^{2}-M^{2})\left(M^{4}-2\epsilon_{1} M^{2}+\epsilon_{2}^{2}\right)}{M^{2}R^{2}-\epsilon_{2}^{2}}}.
\label{totalangularmomentum}\ee

Under the transformation $1\leftrightarrow 2$, the total angular momentum changes its sign, i.e., $J=-J_{(1\leftrightarrow2)}$. This fact means that Eq.(\ref{4parametricsolution}) is indeed a solution for the case of two unequal counter--rotating black holes (with $J_{1}>0$ and $J_{2}<0$). On the other hand, from the energy-momentum tensor associated with the strut, one obtains the following expression for the interaction force between the black holes \cite{Israel,Weinstein}:
\be  \mathcal{F}=\frac{1}{4}(e^{-\gamma_{0}}-1)=\frac{M^{4}-(\sigma^{2}_{1}-\sigma^{2}_{2})^{2}}{4 M^{2}(R^{2}-M^{2})}, \label{Interactionforce}\ee

\noi where $\gamma_{0}$ is the constant value of the metric function $\gamma$ evaluated on the
region corresponding of the strut. It is worthwhile to mention that the interaction force between two
identical counter-rotating black holes ($M_{1}=M_{2}=m$, \, $J_{1}=-J_{2}=j$) in the non--extreme case:
$M=2m$, $\sigma_{1}=\sigma_{2}=\sigma$, as well as in the extreme case:
$M=2m$, $\sigma_{1}=\sigma_{2}=0$, has the same following expression \cite{Varzugin,MRRS,Maria}:
\be  \mathcal{F}=\frac{m^{2}}{R^{2}-4m^{2}}, \qquad R>2m.\label{Interactionforceequal}\ee

Moreover, in the absence of rotation: $J_{1}=J_{2}=0$, $\sigma_{1}=M_{1}$, $\sigma_{2}=M_{2}$,
and $ M=M_{1}+M_{2}$, we recover the well-known expression for the interaction force between
two Schwarzschild black holes \cite{BachWeyl,Weinstein}
\be  \mathcal{F}=\frac{M_{1}M_{2}}{R^{2}-(M_{1}+M_{2})^{2}}, \qquad R>M_{1}+M_{2}, \label{forceSCH}\ee

\noi where $M_{1}$, $M_{2}$ and $R$ are arbitrary and independent parameters.

\section{THE PHYSICAL PARAMETRIZATION}
The relation between the quantities $\sigma_{1}$, $\sigma_{2}$ and the physical Komar parameters of the system can be obtained by means of the Tomimatsu's formulae Eqs.(\ref{Tomi}). Let us use the following simplified form of them  \cite{Tomimatsu0,DietzHoenselaers}:
\be M_{i}=\frac{\omega_{i}}{4}[\Psi|_{\rho=0,z=\alpha_{2i}} -\Psi|_{\rho=0,z=\alpha_{2i-1}}], \qquad
 J_{i}=\frac{\omega_{i}}{2}( M_{i}-\sigma_{i}), \qquad i=1,2\, , \label{mj}\ee
\noi where $\omega_{i}$ are the constant values of the corresponding metric function $\omega$ evaluated over the horizon of each constituent black hole.

The horizons are defined as null hypersurfaces: $\rho=0,-\sigma_{1}\leq z-R/2\leq\sigma_{1}$ and $\rho=0,-\sigma_{2}\leq z+R/2 \leq\sigma_{2}$, separated by a massless strut. A straightforward calculation leads us to the following system of equations for the individual masses and angular momenta of the black holes:
\be  M_{1}=\frac{M^{2}+\epsilon_{2}}{2M},\qquad
M_{2}=\frac{M^{2}-\epsilon_{2}}{2M},\label{themasses}\ee
\be J_{1}= \frac{M_{1}}{2M}\sqrt{\frac{(R+M)(MR-\epsilon_{2})\left(M^{4}-2\epsilon_{1} M^{2}+\epsilon_{2}^{2}\right)}{(R-M)(MR+\epsilon_{2})}},\quad
J_{2}= -\frac{M_{2}}{2M}\sqrt{\frac{(R+M)(MR+\epsilon_{2})\left(M^{4}-2\epsilon_{1} M^{2}+\epsilon_{2}^{2}\right)}{(R-M)(MR-\epsilon_{2})}}. \label{Mj} \ee

From Eq.(\ref{themasses}), it is easy to see that the total mass $M=M_{1}+M_{2}$. However, besides this relation, one obtains the additional relation
\be \sigma_{1}^{2}- \sigma_{2}^{2}=M_{1}^{2}-M_{2}^{2},\label{relation}\ee
replacing Eq.(\ref{relation}) into Eq.(\ref{Mj}) leads to the following expressions for $\sigma_{i}$:
\be\sigma_{1}=\sqrt{M^{2}_{1}-\frac{J^{2}_{1}}{M^{2}_{1}}\frac{(R-M_{2})^{2}-M_{1}^{2}}{(R+M_{2})^{2}-M_{1}^{2}}},
\qquad \sigma_{2}= \sqrt{M^{2}_{2}-\frac{J^{2}_{2}}{M^{2}_{2}}\frac{(R-M_{1})^{2}-M_{2}^{2}}
{(R+M_{1})^{2}-M_{2}^{2}}} . \label{irreduciblemasses}\ee
Eq.(\ref{relation}) implies the following relation between the five physical parameters:
\be J_{1} + J_{2}+ R\left(\frac{J_{1}}{M_{1}}+ \frac{J_{2}}{M_{2}} \right)- M_{1}M_{2}\left(\frac{J_{1}}{M_{1}^{2}}+\frac{J_{2}}{M_{2}^{2}}\right)=0. \label{relationmomentum}\ee

This last relation generalizes the two assumptions (\ref{Bonnorconditions}) made by Bonnor \cite{Bonnor1, Bonnor2}, in order to remove the contribution arising from the massless spinning rods outside the sources, located on the upper and lower part of the symmetry axis, as well as the finite massless spinning rod between the two constituents. Note that Eq.(\ref{totalangularmomentum}) accounts \emph{exactly} for the total angular momentum $J$ as the sum of the individual angular momenta of both constituent black holes, i.e.,
\be J= J_{1}+J_{2}.\ee

Hence, Eq.(\ref{Interactionforce}) reduces to the simple interaction force between two Schwarzschild black holes, where, contrary to what happens in the static case, the distance $R$ is given in terms of the masses and angular momenta of the constituents as follows:
\be R=M_{1}+M_{2}-2\left( \frac{J_{1}+J_{2}}{J_{1}/M_{1} + J_{2}/M_{2}}\right),\label{distance}\ee

\noi therefore the interaction force reduces to:
\be  \mathcal{F}=\frac{M_{1}M_{2}}{R^{2}-(M_{1}+M_{2})^{2}}= -\frac{(J_{1}/M_{1} + J_{2}/M_{2})^{2}}
{4J (J_{1}/M_{1}^{2} + J_{2}/M_{2}^{2})}, \qquad R>M_{1}+M_{2}.\label{newforce}\ee

Moreover, Eq.(\ref{relationmomentum}) implies a dynamical situation between the black holes, since the rotation parameter of one of them is affected by the presence of the other one, according to the relation:
\be J_{2}=-\frac{J_{1}M_{2}}{M_{1}}\left(\frac{R+M_{1}-M_{2}}{R-M_{1}+M_{2}}\right), \qquad
J_{1}=J_{2 (1\leftrightarrow2)}.\ee

On the other hand, if the condition (\ref{relationmomentum}) between the five parameters is not fulfilled, they become independent parameters and a proper contribution of the spin-spin interaction appears in the expression for the interaction force \cite{DietzHoenselaers}.

By using Eq.(\ref{relationmomentum}), one reduces $\sigma_{1}$ and $\sigma_{2}$ to the following expressions:
\be \sigma_{1}=\sqrt{M^{2}_{1}-\frac{J_{1}J_{2}(J_{1}+J_{2})}{M^{2}_{2}J_{1}+M^{2}_{1}J_{2}}},\qquad
\sigma_{2}=\sqrt{M^{2}_{2}-\frac{J_{1}J_{2}(J_{1}+J_{2})}{M^{2}_{2}J_{1}+M^{2}_{1}J_{2}}}.\label{newirreduciblemasses} \ee

Since $\sigma_{1}^{2}>0$ and $\sigma_{2}^{2}>0$, our binary system is composed by black holes (subextreme sources), the interaction force between them takes positive values if $R>M_{1}+M_{2}$, and it implies that $J_{1}>|J_{2}|$,   $J_{1}/M_{1}<|J_{2}|/M_{2}$ and $J_{1}/M_{1}^{2}<|J_{2}|/M_{2}^{2}$ or $J_{1}<|J_{2}|$, $J_{1}/M_{1}>|J_{2}|/M_{2}$ and $J_{1}/M_{1}^{2}>|J_{2}|/M_{2}^{2}$. Note that from Eq.(\ref{irreduciblemasses}) one can recover the case of one isolate black hole by taking the limit $R \rightarrow\infty$ or just by setting to zero the physical parameters of the other body.
\begin{table}[ht]
\centering
\caption{Particular numerical values for the 4-parameter subclass of the Double--Kerr problem.}
\begin{tabular}{c c c c c c c c }
\hline \hline
$\sigma_{1}$&$\sigma_{2}$&$M_{1}$ & $M_{2}$ & $J_{1}$ & $J_{2}$ & $R$ &$J$ \\ \hline
  4.987 & 0.934 & 5  & 1 & 3  & -2    & 7.429  & 1 \\
  1.609 & 5.881 & 2  & 6 & 10.888  & -13.999 & 10 & -3.111 \\
  0.681 & 1.861 & 1  & 2 & 2.5   & -3.0    & 4  & -0.5 \\
  1.972 & 1.972 & 2  & 2 & 2  & -2 & 5 & 0 \\
  3 & 1 & 3  & 1 & 2  & -2 & 4 & 0 \\
  \hline \hline
\end{tabular}
\label{table1}
\end{table}
Table \ref{table1} shows in the first three rows different sets of numerical values for the masses and for the angular momenta of the black holes; the angular momentum of each component having opposite sign. The fourth row displays the case of two equal counter-rotating black holes. The fifth row corresponds to the static case in which the total angular momentum of the system vanishes, i.e., $J=0$. In this case, the horizons of the two black holes can reach each other and the system evolves into one Schwarzschild black hole.

Thus, the expressions for ${\cal{E}}$, $f$, $\omega$ and $\gamma$, describing our 4-parametric solution for two unequal counter-rotating black holes in terms of physical Komar parameters read
\bea {\cal{E}}&=&\frac{\Lambda+2\Gamma}{\Lambda-2\Gamma},\quad f=\frac{|\Lambda|^{2}-4|\Gamma|^{2}}{|\Lambda-2\Gamma|^{2}},
\quad \omega=-\frac{2 {\rm{Im}}\left[(\bar{\Lambda}-2\bar{\Gamma})G\right]}{|\Lambda|^{2}-
4|\Gamma|^{2}},\quad e^{2\gamma}=\frac{|\Lambda|^{2}-4|\Gamma|^{2}}{16\sigma^{2}_{1}\sigma^{2}_{2}
\left[R^{2}-(M_{1}-M_{2})^{2}\right]^{2} r_{1}r_{2}r_{3}r_{4}}, \nonu\\
\Lambda&=&4\sigma_{1}\sigma_{2}M_{1}M_{2}\left(r_{1}r_{2}+r_{3}r_{4}\right)-\mu(r_{1}-r_{2})(r_{3}-r_{4})\nonu\\
 &+& \sigma_{1}\sigma_{2}(R^{2}-M^{2}_{1}-M^{2}_{2})(r_{1}+r_{2})(r_{3}+r_{4})
- i\nu \left[\sigma_{1}(r_{1}+r_{2})(r_{3}-r_{4})-\sigma_{2}(r_{1}-r_{2})(r_{3}+r_{4})\right],\nonu\\
\Gamma&=&-\sigma_{1}M_{1}\left[\sigma_{2}(R^{2}-M^{2}_{1}+M^{2}_{2})(r_{3}+r_{4})
+2M^{2}_{2}R(r_{3}-r_{4})\right] \nonu\\
&-& \sigma_{2}M_{2}\left[\sigma_{1}(R^{2}+M^{2}_{1}-M^{2}_{2})(r_{1}+r_{2})-2M^{2}_{1}R(r_{1}-r_{2})\right]
+ i\nu\left[\sigma_{1}M_{1}\left(r_{3}-r_{4}\right)-\sigma_{2}M_{2}\left(r_{1}-r_{2}\right)\right], \nonu\\
G&=&2z\Gamma +4\sigma_{1}\sigma_{2}M_{1}M_{2}R\left(r_{1}r_{2}-r_{3}r_{4}\right)
+\sigma_{1}M^{2}_{2}(R^{2}+M^{2}_{1}-M^{2}_{2})(r_{1}+r_{2})(r_{3}-r_{4})\nonu\\
&+& \sigma_{2}M^{2}_{1}(R^{2}-M^{2}_{1}+M^{2}_{2})(r_{1}-r_{2})(r_{3}+r_{4})
- i\nu(M^{2}_{1}-M^{2}_{2})\left(r_{1}-r_{2}\right)(r_{3}-r_{4})\label{Fourparametricsolution}\\
&-& \sigma_{1}M_{1}\left\{\sigma_{2}R(M^{2}_{1}+3M^{2}_{2}-R^{2})(r_{3}+r_{4})
+2\left[\mu + (\sigma^{2}_{1}+\sigma^{2}_{2})M^{2}_{2}\right](r_{3}-r_{4})\right\}\nonu\\
&+& \sigma_{2}M_{2}\left\{\sigma_{1}R(3M^{2}_{1}+M^{2}_{2}-R^{2})(r_{1}+r_{2})-
2\left[\mu + (\sigma^{2}_{1}+\sigma^{2}_{2})M^{2}_{1}\right](r_{1}-r_{2})\right\}\nonu\\
&-&  i \nu \left\{ \sigma_{1}M_{1}\left[R(r_{3}-r_{4})-2\sigma_{2}(r_{3}+r_{4}) \right]+
\sigma_{2}M_{2}\left[R(r_{1}-r_{2})+2\sigma_{1}(r_{1}+r_{2}) \right]\right\},\nonu\\
\mu &:=&(1/2) \left[(\sigma^{2}_{1}+\sigma^{2}_{2})(R^{2}-M^{2}_{1}-M^{2}_{2}) -(M^{2}_{1}+M^{2}_{2})R^{2} +(M^{2}_{1}-M^{2}_{2})^{2}\right],  \nonu\\
\nu &:=&(1/\sqrt{2})\left(R-M_{1}-M_{2} \right) \sqrt{\frac{J^{2}_{1}}{M_{1}^{2}}\left(R+M_{1}-M_{2}\right)^{2}+\frac{J^{2}_{2}}{M_{2}^{2}}\left(R-M_{1}+M_{2}\right)^{2}} \nonu.\eea

The Ernst potential on the symmetry axis reads
\bea  e(z)&=&\frac{e_{+}}{e_{-}}, \nonu \\
e_{\pm}&=&z^{2} \mp M z -\left(\frac{R}{2} \pm M_{1}\right)\left(\frac{R}{2} \mp M_{2}\right) + \left(\frac{R^{2}-M^{2}}{4}\right)F^{\pm 1}  \\
&-&\frac{R-M}{R+M}\left[ \left(\frac{R+M}{2}\right)F^{\pm 1/2}+\frac{i}{\sqrt{2}} \sqrt{\frac{J^{2}_{1}}{M_{1}^{2}}F +\frac{J^{2}_{2}}{M_{2}^{2}}F^{-1}} \right]^{2},\nonu\\
F&:=&\frac{R+M_{1}-M_{2}}{R-M_{1}+M_{2}}, \nonu \eea

\noi where $\sigma_{i}$ and $r_{n}$ are given by Eq.(\ref{irreduciblemasses}) and Eq.(\ref{thedistances}), respectively. Besides the five physical parameters satisfy the generalized Bonnor condition Eq.(\ref{relationmomentum}).

\subsection{Thermodynamical properties}
For each component of the binary system, the Smarr formula for the mass \cite{Smarr} holds, i.e.,
\be M_{i}=\frac{\kappa_{i}S_{i}}{4\pi}+2\Omega_{i}J_{i}= \sigma_{i}+2\Omega_{i}J_{i},
\quad i=1,2,\label{Smarrformula}\ee

\noi where $\kappa_{i}$ is the surface gravity, $S_{i}$ is the area of the horizon, $\Omega_{i}$
the angular velocity and $J_{i}$ the angular momentum for each constituent black hole. Notice that this last formula implies that $M_{i}>\sigma_{i}$. In order to calculate the values of $\kappa_{i}$ and $\Omega_{i}$, one can use the following relations \cite{Tomimatsu1, DietzHoenselaers}:
\be \kappa_{i}=\sqrt{-\omega^{-2}_{i}e^{-2\gamma_{i}}},  \qquad \Omega_{i}=\omega^{-1}_{i},\ee

\noi being $\omega_{i}$ and $\gamma_{i}$ the constant values of the corresponding metric functions $\omega$ and $\gamma$ evaluated over the horizon of each constituent, while $e^{2\gamma}$ is negative at the horizon \cite{DietzHoenselaers}. By means of the solution Eq.(\ref{Fourparametricsolution}) and Eq.(\ref{relationmomentum}), it is straightforward to obtain the following expressions for the angular velocities $\Omega_{i}$, the surface gravities $\kappa_{i}$, and the area of the horizons $S_{i}$:
\bea \begin{split} \Omega_{1}&=\frac{J_{1}[(R-M_{2})^{2}-M_{1}^{2}]}{2M_{1}^{2}(M_{1}+\sigma_{1})
[(R+M_{2})^{2}-M_{1}^{2}]}=\frac{J_{2}(J_{1}+J_{2})}{2(M_{1}+\sigma_{1})(M_{2}^{2}J_{1} + M_{1}^{2}J_{2})}, \\
\kappa_{1}&= \frac{\sigma_{1}(R+M_{1}-M_{2})}{2M_{1}(M_{1}+\sigma_{1})(R+M)}=
\frac{\sigma_{1}(M_{1}-M_{2})J_{2}}{2(M_{1}+\sigma_{1})(M_{2}^{2}J_{1}+M_{1}^{2}J_{2})},\\
S_{1}&=\frac{8\pi M_{1}(M_{1}+\sigma_{1})(R+M)}{(R+M_{1}-M_{2})}=\frac{8\pi (M_{1}+\sigma_{1})(M_{1}^{2}J_{2}+M_{2}^{2}J_{1})}{(M_{1}-M_{2})J_{2}},\\
\Omega_{2}&=\Omega_{1(1\leftrightarrow2)}, \quad \kappa_{2}=\kappa_{1(1\leftrightarrow2)},
\quad S_{2}=S_{1(1\leftrightarrow2)}.\end{split}\eea

\noi  In the limit when the sources are far away from each other, the angular velocities reduce to
\bea \begin{split}
\Omega_{i}&=\frac{J_{i}F_{i}}{2M_{i}^{2}(M_{i}+\sigma_{i})}, \\
F_{1}&\simeq 1-\frac{4M_{2}}{R}+\frac{8M^{2}_{2}}{R^{2}}-\frac{4M_{2}(M^{2}_{1}+3M^{2}_{2})}{R^{3}}+
O\left(\frac{1}{R^{4}}\right),  \\
F_{2}&\simeq 1-\frac{4M_{1}}{R}+\frac{8M^{2}_{1}}{R^{2}}-\frac{4M_{1}(M^{2}_{2}+3M^{2}_{1})}{R^{3}}+
O\left(\frac{1}{R^{4}}\right),
\end{split}\eea
notice that the proper contribution to the angular velocity $\Omega_{i}$ coming from the angular momentum $J_{i}$ begins at the third order of the expansion, i.e., $\Omega_{i}\simeq O(1/R^{3})$) \cite{Varzugin}.

Additionally, in the limit $M_{1}=M_{2}=m$,\, $\sigma_{1}=\sigma_{2}=\sigma$ and $J_{1}=-J_{2}=j$, our solution reduces to the one for the case of two identical counter-rotating black holes. The unique $\sigma$ reads
\be \sigma=\sqrt{m^{2}-\frac{j^{2}}{m^{2}}\left(\frac{R-2m}{R+2m}\right)}. \label{uniquesigma}\ee

Therefore, there does not exist a dynamic scenario between the black holes, since relation  Eq.(\ref{relationmomentum}) is satisfied and the parameters $m$, $j$ and $R$ become independent. This particular
case belongs to a 3-parameter subclass of the DKN solution \cite{KramerNeugebauer}, where the total angular momentum of the system vanishes, i.e., $J=0$.

\subsection{Singularities off the axis}
\begin{figure}[ht]
\begin{minipage}{0.49\linewidth}
\centering
\includegraphics[width=5cm,height=5cm]{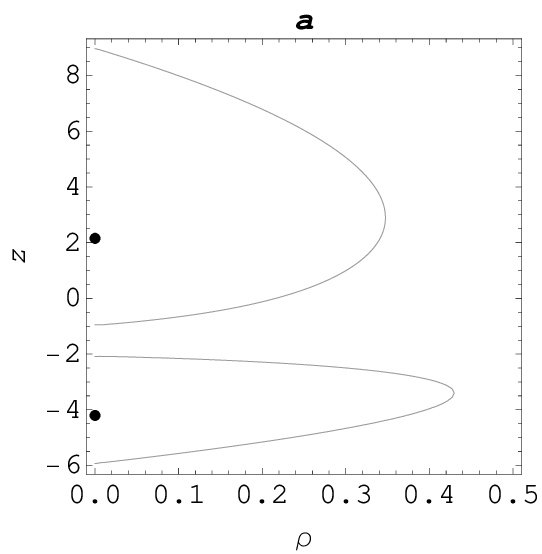}
\end{minipage}
\begin{minipage}{0.49\linewidth}
\centering
\includegraphics[width=5cm,height=5cm]{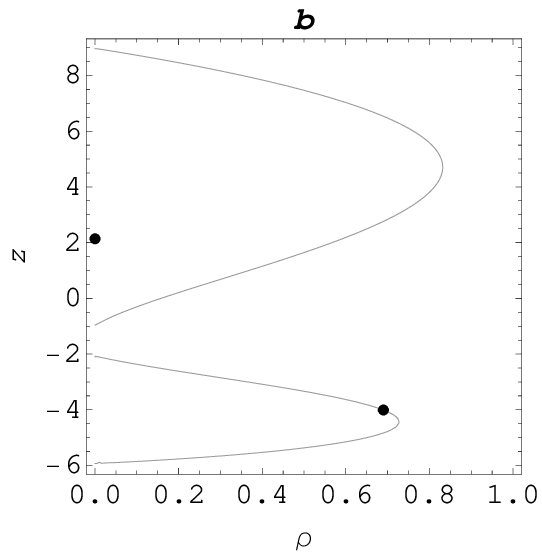}
\end{minipage}
\caption{(a) For positive masses, there exists no singularity off the axis, for the values: $\sigma_{1}=4.973$, $\sigma_{2}=1.931$, $M_{1}=5$, $M_{2}=2$, $J_{1}=6.82$, $J_{2}=-6$ and $R=8$; (b) If one of the masses is negative, there appears a ring singularity off the axis and the system becomes co-rotating, for the values: $\sigma_{1}=4.973$, $\sigma_{2}=1.931$, $M_{1}=5$, $M_{2}=-2$, $J_{1}=1.0$, $J_{2}=6$ and $R=8$. The ring singularity is located at $\rho\simeq0.69$, $z\simeq-4.01$.}
\label{Stationarysurfaces}\end{figure}
Since $M_{i}>0$, Eq.(\ref{Fourparametricsolution}) describes a binary system composed by two unequal counter-rotating black holes separated by a massless strut, whose respective interior naked singularities lie on the symmetry axis, in the region corresponding to their horizons. Nevertheless, if one of the masses is negative, even when the total ADM mass is positive \cite{ADM}, the solution characterized by Eq.(\ref{Fourparametricsolution}) presents ring singularities off the axis and the system turns out to be co-rotating instead of counter-rotating (see Eq.(\ref{Mj})). For instance, if $M_{2}<0$, Eq.(\ref{Fourparametricsolution}) describes a system composed by a black hole and a naked singularity (ring singularity off the axis).

By setting $f=0$, this fact can be observed in the stationary limit surfaces of Fig. \ref{Stationarysurfaces}. The location of such ring singularity off the axis can be calculated as one root of the denominator of Ernst potential Eq.(\ref{Fourparametricsolution}).

\section{CONCLUDING REMARKS}
In this work, we present an exact solution describing a binary system constituted by two unequal counter-rotating black holes with a massless strut in between. We derive a 4-parameter subclass involving a simple algebraic relation
between the five physical parameters. This relation generalizes, for systems of unequal black holes, the two assumptions made by Bonnor \cite{Bonnor1,Bonnor2} in order to avoid the contribution from the massless spinning rods outside the black holes and it defines a dynamic scenario between the two black holes, for which the physical and geometrical properties of one black hole are affected by the presence of the other one. Therefore, the interaction force provided by the strut results to be of the same form as the Schwarzschild type, where the coordinate distance becomes a function of physical Komar masses and angular momenta. This solution reduces to the one for the case of identical constituents \cite{MRRS}.\nl

On the other hand, in the extreme limit: $\sigma_{1}=0$ and $\sigma_{2}=0$, the unequal and opposite angular momenta per unit mass, in absolute value, are greater than their corresponding positive masses: $|J_{i}|/M_{i}>M_{i}>0$,  according to the expressions
\be \frac{J_{1}}{M_{1}}=\epsilon M_{1}\sqrt{\frac{(R+M_{2})^{2}-M_{1}^{2}}{(R-M_{2})^{2}-M_{1}^{2}}}, \qquad \frac{J_{2}}{M_{2}} =-\epsilon M_{2}\sqrt{\frac{(R+M_{1})^{2}-M_{2}^{2}}{(R-M_{1})^{2}-M_{2}^{2}}}, \qquad \epsilon=\pm1.\label{extremerelation}\ee

However, in this particular case, the condition established between the five parameters is satisfied only if both constituent black holes are equal: $M_{1}=M_{2}=m$ and $J_{1}=-J_{2}=j$. Therefore, the total angular momentum of the system vanishes, and there exists a stable distance $R$ in which the extremality condition is achieved \cite{MRRS}:
\be R =\frac{2m (j^{2}+m^{4})}{j^{2}-m^{4}}>2m, \qquad |j|/m>m>0.\label{distance}\ee

The property of the angular momentum per unit mass exceeding the value of the mass in the identical case, was first pointed out by Herdeiro et al.\cite{Herdeiro}, and it can be obtained also as a trivial consequence of the work of Varzugin \cite{Varzugin}.

It is worthwhile to mention that even when the interaction force for identical constituents, has the same form in the extreme case as well as in the non-extreme case, it does not mean, that the force can take the same value. In fact Eq.(\ref{uniquesigma}) can be written in the following form:
\be \sigma=\sqrt{\left(\mathcal{F}-\frac{j^{2}}{m^{2}(R+2m)^{2}}\right)(R^{2}-4m^{2})}, \qquad \mathcal{F}=\frac{m^{2}}{R^{2}-4m^{2}}, \label{uniquesigma2}\ee

\noi and it implies that the force can take positive values given by
\be \mathcal{F} \geq \frac{j^{2}}{m^{2}(R+2m)^{2}}>0, \qquad R>2m. \label{newforce} \ee

The equality is reached when the black holes become extreme and the coordinate distance takes the particular value given in Eq.(\ref{distance}) and therefore the interaction force reduces to \cite{Maria}
\be \mathcal{F} =\frac{(j^{2}-m^{4})^{2}}{16m^{4}j^{2}}, \qquad |j|/m>m>0. \label{newforceextreme} \ee

\noi for such value of the force, one can prove the equality into the geometrical inequality between extreme black holes with struts, provided by Clement \cite{Maria}
\be \sqrt{1+4\mathcal{F}}=\frac{8\pi |j|}{S_{ext}}, \qquad S_{ext}=\frac{16\pi m^{2}j^{2}}{j^{2}+m^{4}}.\ee

\noi where $S_{ext}$ is the area of the horizon of the extreme black hole. It is important to mention that Eq.(\ref{extremerelation}), relating masses and angular momenta, is of the same form as the relations presented in \cite{Cabrera}, in the context of a binary system constituted by two extreme Reissner--Nordstr\"{o}m black holes with a strut in between.\nl

It is well-known that in the case of two extreme balancing constituents (in the absence of strut), at least one of the two bodies is endowed with negative mass and consequently it cannot be a black hole. Instead of being a black hole it turns out to be a naked singularity (ring singularity off the axis), making singular the Ernst potential and the entire solution is not regular outside the symmetry axis. Ring singularities off the axis in the framework of the double-Kerr solution have been always associated exclusively with a negative mass of one of the constituents.

On the other hand, Manko \emph{et al} \cite{MankoRuizSadovnikova} consider the interaction force associated with the strut and the entire metric for the case of non-identical extreme black holes, in the counter-rotating case as well as in the co-rotating case. Since in our case, the two bodies tend to be identical for some ``stable distance" and the force takes a particular positive value defined by Eq.(\ref{newforceextreme}), \cite{MankoRuizSadovnikova} deals with a more general situation than our extreme solution (recovered if $q=0$ in Eq.(9) of \cite{MankoRuizSadovnikova}).

However, we believe that as it happens in the vacuum case, also in the electrovacuum case, there will always exist a stable distance, which can be modified by the presence of the electric charge. This fact can be observed in Eq.(58) of \cite{MankoRuizSadovnikova}. The issue of a massless ring singularity off the symmetry axis emerging in certain binary black hole systems is intriguing and deserves further investigation.

The technical detail for removing the NUT sources outside the two rotating black holes is not a trivial
problem and it restricts the possibilities for finding exact solutions to more general problems related to the counter/co-rotating cases. Finally, it would be interesting to provide electric charge to such configurations, since the charge could avoid the presence of the conical singularity in between. These issues remain as future works to analyze.

\section*{ACKNOWLEDGEMENTS}
We would like to thank E. Ruiz, V. S. Manko, V. Perlick and N. G\"{u}rlebeck for useful discussions and literature hints. This research was supported by DFG--CONACyT Grant No. B330/418/11, by CONACyT Grant No. 166041F3, by Deutscher Akademischer Austausch Dienst (DAAD) Fellowship No. A/10/77743, and by CONACyT Fellowship with CVU No. 173252. CL acknowledges also support by the DFG Research Training Group 1620 ``Models of Gravity'', and by the center of excellence QUEST.

\end{document}